\documentclass{osa-article}

\definecolor{authorcolor}{RGB}{20,120,182}
\definecolor{urlblue}{RGB}{46,46,177}
\usepackage{graphics}      
\usepackage{graphicx}      
\usepackage{longtable}     
\usepackage{url}           
\usepackage{bm}            
\usepackage{braket}
\usepackage{bbm}
\usepackage{color}
\usepackage{comment}
\articletype{Research Article}

\begin{document}


\title{Theoretical analysis of glide-$Z_2$ magnetic topological photonic crystals}
\author{Heejae Kim,\authormark{1} Hengbin Cheng,\authormark{2} Ling Lu,\authormark{2}, and Shuichi Murakami\authormark{1,3,*}}

\address{\authormark{1}Department of Physics, Tokyo Institute of Technology, 2-12-1 Ookayama, Meguro-ku, Tokyo, 152-8551, Japan\\
\authormark{2}Institute of Physics, Chinese Academy of Sciences/Beijing National Laboratory for Condensed Matter Physics, Beijing 100190, China\\
\authormark{3}TIES, Tokyo Institute of Technology, 2-12-1 Ookayama, Meguro-ku, Tokyo, 152-8551, Japan}

\email{\authormark{*}murakami@stat.phys.titech.ac.jp} 



\begin{abstract}
Gapped systems with glide symmetry can be characterized by a $Z_2$ topological invariant. We study the magnetic photonic crystal with a gap between the second and third lowest bands, which is characterized by
the nontrivial glide-$Z_2$ topological invariant that can be determined by symmetry-based indicators. We show that under the space group No.~$\bm{{\it 230}}$ ($Ia\bar{3}d$), the topological invariant is equal to a half of the number of photonic bands below the gap, 
and therefore, the band gap between the second and third lowest bands is always 
topologically nontrivial, and to realize the topological phase, we need to open a gap for the Dirac point at the $P$ point
by breaking time-reversal symmetry. 
With staggered magnetization, the photonic bands are gapped and the photonic crystal becomes  
 topological, whereas with uniform magnetization, a gap does not open, which 
 can be attributed to the minimal band connectivity exceeding two in this case.
 By introducing the notion of Wyckoff positions,
we show how the topological characteristics are determined from the structure of the photonic crystals.\end{abstract}

\section{Introduction}

\begin{figure}
\centering
\includegraphics[width=10cm,clip]{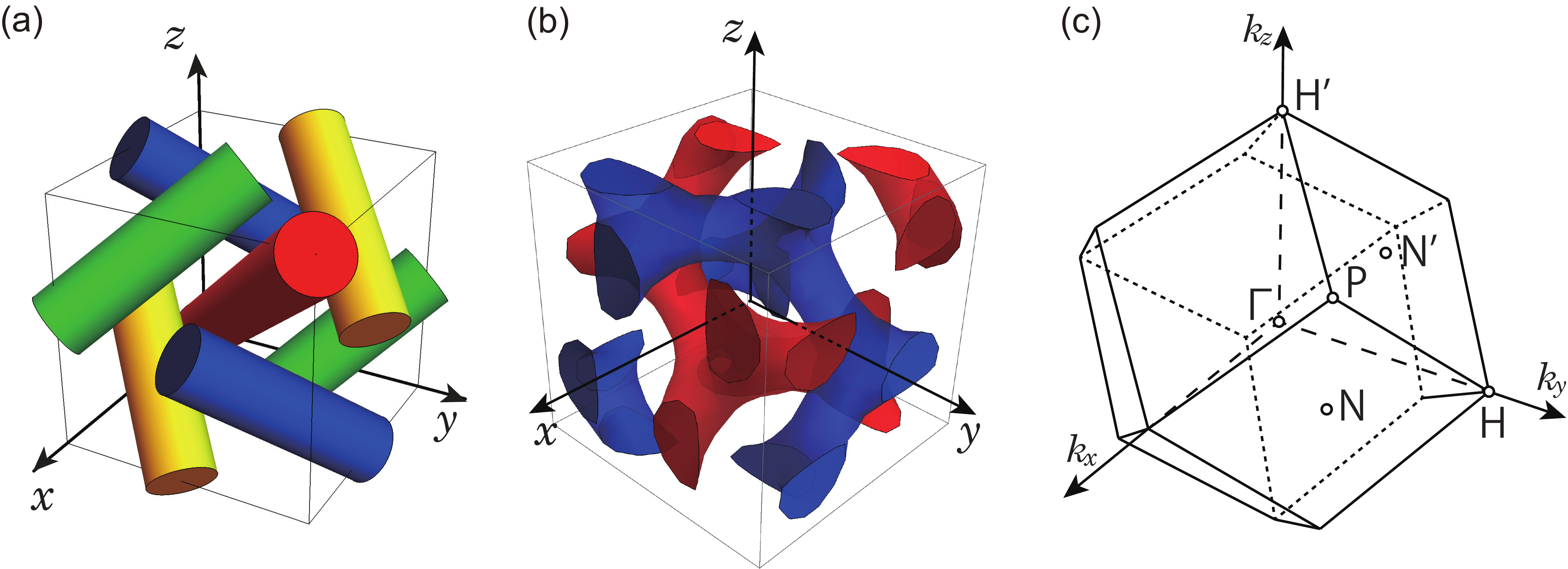}
\caption{The BPI and DG photonic crystals in the body-centered cubic (BCC) lattice and their Brillouin zone (BZ) of the BCC lattice. (a) The BPI photonic crystal, shown in the cubic unit cell of length $a$. It consists of four identical dielectric rods oriented along the BCC lattice vectors along (111) (red), $(11\bar{1})$ (yellow), $(1\bar{1}1)$ (blue) and $(\bar{1}11)$ (green), and they go through the points (0,0,0)$a$, (0,0.5,0)$a$, (0.5,0,0)$a$ and (0,0,0.5)$a$, respectively. (b) The DG photonic crystal whose dielectric regions are given by $g(\bm{r})>\lambda_{\mathrm{iso}} = 1.1$ (blue) and $g(-\bm{r})>\lambda_{\mathrm{iso}} = 1.1$ (red) of Eq.~(\ref{eq:dg_function}). (c) The BZ of the BCC lattice and its high-symmetry points $\Gamma$, $P$,  $N$ and H.}
\label{fig:phc_uc}
\end{figure}

Manipulations of magnetic topological materials are an intriguing and promising topic in condensed matter physics.
In particular, a glide-symmetric topological crystalline insulator (TCI) exhibits a $Z_2$ topological phase in class A
hosting a single surface Dirac cone topological phase with the nontrivial $Z_2$ topopogical invariant 
\cite{Fang2015prb91, Shiozaki2015prb91,Kim2019prb100,Shiozaki2018arXiv180206694}.
One of our targets for material realizations is bosonic systems.
In photonics,
a new type of topological band-crossing points beyond the Weyl and Dirac points emerges.
For example,
at a generalized Dirac point \cite{Lu2016nphys12}, the bands are four-fold degenerate and they split into three or four bands along any direction, in contrast with a Dirac points, which generally splits into two sets of double degenerate bands along any direction.
When the system is perturbed, this generalized Dirac point becomes
line nodes, Weyl points, or opens a band gap \cite{Lu2013nphoton7, Lu2016nphys12}.
As an example, in the photonic crystal having a structure in the first blue phase of liquid crystals (BPI) \cite{Lu2016nphys12} shown in Fig.~\ref{fig:phc_uc},
by introducing magnetization to break time-reversal symmetry (TRS) 
the generalized Dirac point opens a band gap and the system goes into topological phases protected by glide symmetry.
On the other hand,
in the double gyroid (DG) photonic crystal \cite{Lu2013nphoton7} (Fig.~\ref{fig:phc_uc}(b)) belonging to the same space group $\bm{{\it 230}}$ (henceforth, we call a space group by its number in bold italic following in Ref.~\cite{Hahn2002ITA}) with the BPI photonic crystal,
the system has Weyl points \cite{murakami2007phase} between the fourth and fifth lowest bands in the absence of TRS.
Nonetheless, how the band structures and topological properties are determined from the structure of the photonic crystals needs to be clarified.

In the present paper, we show how this topological phase can be manipulated, based on the relationship between space group representations and band structures.
In the BPI photonic crystal, the gap opening at non-equivalent $P$ and $P^\prime$ points in the Brillouin zone (BZ) (Fig.~\ref{fig:phc_uc}(c)) makes the system gapped and topologically nontrivial.
However, the reason why this structure leads to the topolgical phase, and how the eigenmodes at these $P$ and $P'$ points affect the topological phase is not yet clear. 
In this paper, we clarify these points, by using the formula of the glide-$Z_2$ invariant which we derived in our previous paper \cite{kim2020-nonprimitive}. 
In particular, in the space group $\bm{{\it 230}}$, which the photonic crystals 
in the previous works \cite{Lu2013nphoton7, Lu2016nphys12} belong to, the topological invariant is equal to a half of the number of bands below the gap in terms of modulo 2, 
and therefore, the gap between the second and third lowest bands is always topologically nontrivial. On the other hand, 
in the presence of the TRS, since all modes at the $P$ points are fourfold degenerate, 
the band gap between the second and the third lowest modes is closed at the $P$ point. Therefore, we conclude that 
when this gap is open by breaking the TRS, 
  the photonic crystal is a topological one protected by the glide symmetry,
  as has been realized  in the BPI photonic crystal \cite{Lu2016nphys12} with staggered magnetization.
We also show that it automatically becomes a higher-order topological insulators \cite{Song2017prl119,Schindler2018nphys14,Benalcazar2017sci357,Benalcazar2017prb96,
Langbehn2017prl119,Schindler2018sciadv4}.  On the other hand, in a uniform magnetization, a gap does not open between the second and third lowest bands, which 
is explained by the notion of the minimal band connectivity \cite{Watanabe2018prl121}. Then, we propose a way to design such a topological photonic crystal by introducing the notion of Wyckoff positions into the design of photonic crystals.

This paper is organized as follows. Section~\ref{sec:prev} is devoted to reviewing the previous works necessary for the discussion of the main part of this paper.
In Sec.~\ref{sec:symm}, we analyze the irreducible representations of the photonic crystals considered, and discuss which cases will realize topological photonic crystals. 
In Sec.~\ref{sec:manipulation} we introduce the notion of the Wyckoff positions into the design of the photonic crystals, and discuss 
the relationship between the positions of the dielectrics and topological phases.
We conclude the paper in Sec.~\ref{sec:conclusion}.

\section{$Z_2$ topological photonic crystal with glide symmetry}
\label{sec:prev}
In this section, we first review the BPI photonic crystal, which is proposed to be in the topological phase with glide symmetry in Ref.~\cite{PhysRevResearch.2.013300},
and then we analyze its characteristics from our viewpoint. 

\subsection{Review on the BPI topological photonic crystal}
We consider a 3D photonic crystal 
in a body-centered cubic (BCC) lattice,
composed of four identical dielectric rods.
A schematic illustration of the system is shown in Fig.~\ref{fig:phc_uc}(a). We take the $x$, $y$, $z$  axes to be parallel to the edges of the cubic unit
cell. The four dielectric rods are along the BCC primitive vectors along (111), $(11\bar{1})$, $(1\bar{1}1)$ and $(\bar{1}11)$, and they go through the points (0,0,0)$a$, (0,0.5,0)$a$, (0.5,0,0)$a$ and (0,0,0.5)$a$, respectively,  where $a$ is the lattice constant of the cubic cell.
The space group is $\bm{{\it 230}}$ ($Ia\bar{3}d$), which is nonsymmorphic, containing glide reflections and inversion.

\begin{figure}
\centering
\includegraphics[scale=0.24]{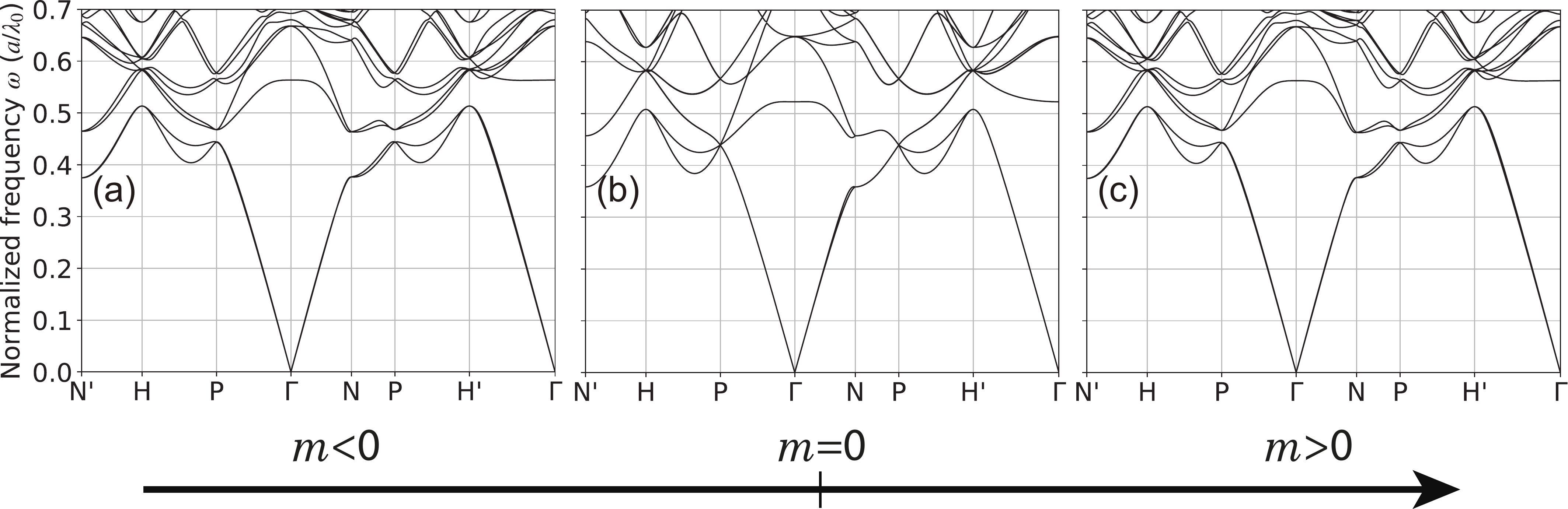}
\caption{Band structures for (a)-(c) the BPI photonic crystal. The band structures with magnetization where (a) $\varepsilon_{33} = 11$ and $\kappa=10i, 10i, -10i, -10i$ and (c) with $\varepsilon_{33} = 11$ and $\kappa = -10i, -10i, 10i, 10i$ for the red, yellow, green and blue rods shown in Fig.~\ref{fig:phc_uc}(a), respectively, are exactly the same, because of the TRS, and both sides are topological.}
\label{fig:both_side}
\end{figure}

When we set the dielectric constant of the rods to be $\varepsilon = 11$ and the radius of the rods to be $r = 0.13a$,
the fourfold generalized Dirac point appears at two non-equivalent $P$ points in the BZ formed by the bands 1, 2, 3 and 4 counted from the lowest band (Fig.~\ref{fig:both_side}(b)),  because of the cubic and time-reversal (TR) symmetries \cite{Lu2016nphys12}.
If the magnetization is present in these rods with the glide symmetry preserved,
the band structure acquires a band gap between bands 2 and 3 by breaking the TRS and it becomes the topological phase with the nontrivial $Z_2$ topopogical invariant 
\cite{Fang2015prb91, Shiozaki2015prb91,Kim2019prb100}
protected by glide symmetry 
\cite{Lu2016nphys12}.
To be more specific, 
in the dielectric tensor of the rod $\bm{\varepsilon}$, we add off-diagonal imaginary terms $\kappa$, which breaks TRS,
\begin{equation}
\bm{\varepsilon} =
\begin{pmatrix}
\varepsilon_\parallel & \kappa & 0 \\ -\kappa & \varepsilon_\parallel & 0 \\ 0 & 0 & \varepsilon_{33}
\end{pmatrix} ,
\label{eq:mag_epsilon}
\end{equation}
where $\varepsilon_\parallel\ (>0)$ and $\varepsilon_{33}\ (>0)$ are dielectric constants with 
 $\varepsilon_\parallel^2 - |\kappa|^2 = \varepsilon_{33}^2$ \cite{Lu2016nphys12, Lu2013nphoton7}, and 
 $\kappa$ is a non-zero imaginary number. 
These off-diagonal imaginary terms lead to gyroelectric response of materials \cite{Haldane2008prl100}. Similarly, gyromagnetic response in ferrites is generated by off-diagonal terms of the permeability tensor \cite{Wang2009nature461}.
In the calculation, 
the off-diagonal terms are introduced in a staggered way as $\kappa=\pm 10i, \pm 10i, \mp 10i, \mp10i$ with $\varepsilon_{33} = 11$ for the dielectric rods 
along (111) (red), $(11\bar{1})$ (yellow), $(1\bar{1}1)$ (blue) and $(\bar{1}11)$ (green) shown in Fig.~\ref{fig:phc_uc}(a).
In Fig.~\ref{fig:both_side}(a) and (c), we show the band structures of the BPI photonic crystal with these two patterns of the sign of $\kappa$, with the 
rod along the (111) direction  having $\kappa_{(111)} = \pm 10i$, respectively.
This value of $\kappa$ does not reflect that in real materials, but is only for demonstration of the behaviors of the gap.
At $\kappa_{(111)}=0$ preserving TRS, a four-fold degenerate generalized Dirac point exists at the $P$ point (Fig.~\ref{fig:both_side}(b)).
Then by breaking the TRS via nonzero $\kappa_{(111)}$, a band gap opens between the bands 2 and 3 at the $P$ points, and 
both of these cases with $\kappa_{(111)}=\pm 10i$ are found to be topological in Ref.~\cite{Lu2016nphys12}.

\subsection{Discussion on symmetry and topology of the topological photonic crystal}
Thus, regardless of the sign of $\kappa_{(111)}$, the band gap opens at the $P$ point and this gap becomes 
topological. This is related with the fact that the irreducible representations (irreps) at the $P$ points do not affect the values of the 
topological invariant; namely, the gap at $P$ is inverted between $\kappa_{(111)} = +10i$ and $\kappa_{(111)} = -10i$,
but it does not affect the topological invariant.
This situation is similar to the Kane-Mele model, which is a well-known model for the two-dimensional topological insulator  (TI) phase with 
TRS \cite{Kane2005prl95-1, Kane2005prl95-2}. The Kane-Mele model is topological, 
 regardless of the sign of the spin-orbit coupling $\lambda_{\mathrm{so}}$, which opens the gap at the $K$ point. 
Thus both in the BPI photonic crystal and in the Kane-Mele model, the topological invariant is independent of the irreps at $P$ or $K$ points, and regardless of the sign of the parameter ($\kappa_{(111)}$ or $\lambda_{\mathrm{so}}$) to open a gap, the system becomes topological as long as the parameter is nonzero. More details are presented in Sec. S1 in the Supplemental Document.

\section{Symmetry consideration for band theory of glide-symmetric $Z_2$ magnetic topological photonic crystal }
\label{sec:symm}

In this section, we discuss what determines the topological invariant in the photonic crystal from the symmetry viewpoint.
We first investigate the irreps for the glide-symmetric $Z_2$ magnetic topological photonic crystals, i.e. the BPI 
\cite{Lu2016nphys12}, and DG \cite{Lu2013nphoton7}
photonic crystals.
Then, we calculate the glide-$Z_2$ invariant by using our new formula derived in Ref.~\cite{kim2020-nonprimitive}.

\subsection{Representations at the high-symmetry points from symmetry considerations}
\label{subsec:rep_HSP}

We first explain irreps at high-symmetry points in the BZ in the space group No.~$\bm{{\it 230}}$.
There are four kinds of high-symmetry points $\Gamma$, $H$, $N$, and $P$ in the BZ (see Fig.~\ref{fig:bz7-15} (c)).
Among these four high-symmetry points, the points $\Gamma$, $H$, and $N$ are TRIMs in BZ, but $P$ is not.
Even though $P$ is not a TRIM,
the bands at the $P$ point should be degenerate in the presence of TRS.
This double degeneracy comes from combined symmetry $\mathcal{T}^\prime = \mathcal{T} G_y$ where $\mathcal{T}$ is time-reversal operation, 
and $G_y$ is the glide symmetry $G_y=\{M_y|\frac{a}{2}\hat{z}\}$,
where
$\hat{\bm{x}}, \hat{\bm{y}}$ and $\hat{\bm{z}}$ denote the unit vectors along $\bm{x}, \bm{y},$ and $\bm{z}$ directions, respectively.
As $\mathcal{T}^\prime$ is an antiunitary symmetry, and it satisfies $\left( \mathcal{T}^\prime \right)^2 = G_y^2 = T_z = -1$ at the $P$ point,
where $T_z=\{E|a\hat{z}\}$, and $E$ is an identity operation,
the bands are always doubly degenerate at the $P$ point.

Our strategy is the following.
We first start with a vacuum having a homogeneous dielectric constant $\varepsilon = 1$.
Thereby,
the eigenmodes behave as electromagnetic plane waves with a linear polarization in free space.
From symmetry considerations for these plane-wave basis functions,
we can identify representations at high-symmetry points which characterize the eigenmodes at $\varepsilon=1$ (Fig.~\ref{fig:phc_evolution}(a)).
Next, we gradually increase the dielectric constant from $\varepsilon=1$ within the dielectrics to describe a photonic crystal.
Then, the eigenmodes split into irreps at the high-symmetry points in the BZ, from which we discuss the topological invariant.


\begin{table}
\caption{Eigenmodes with plane waves at the $H^{(0)}$ point. $\bm{\hat{x}}, \bm{\hat{y}}$, and $\bm{\hat{z}}$ denote the unit vectors along $\bm{x}, \bm{y}$, and $\bm{z}$ directions, respectively. We
put $\omega=e^{2\pi i/3}$ here.}
$$
\begin{array}{c|c} \hline \hline
H_1 & 
\begin{matrix}
\psi^1_{H_1} = \sin \left( \frac{2\pi}{a} x \right) \bm{\hat{y}} + \sin \left( \frac{2\pi}{a} y \right) \bm{\hat{z}} +\sin \left( \frac{2\pi}{a} z \right) \bm{\hat{x}}  \\
\psi^2_{H_1} = \cos \left( \frac{2\pi}{a}y \right) \bm{\hat{x}} +  \cos \left( \frac{2\pi}{a} x \right) \bm{\hat{z}} +  \cos \left( \frac{2\pi}{a} z \right) \bm{\hat{y}}
\end{matrix}
\\ \hline
H_2 & 
\begin{matrix}
\psi^1_{H_2} = \sin \left( \frac{2\pi}{a} x \right) \bm{\hat{y}} + \omega \sin \left( \frac{2\pi}{a} y \right) \bm{\hat{z}} + \omega^2\sin \left( \frac{2\pi}{a} z \right) \bm{\hat{x}} \\
\psi^2_{H_2} = \cos \left( \frac{2\pi}{a}y \right) \bm{\hat{x}} + \omega\cos \left( \frac{2\pi}{a} x \right) \bm{\hat{z}} + \omega^2 \cos \left( \frac{2\pi}{a} z \right) \bm{\hat{y}}
\end{matrix}
\\ \hline
H_3 & 
\begin{matrix}
\psi^1_{H_3} = \sin \left( \frac{2\pi}{a} x \right) \bm{\hat{y}} + \omega^2 \sin \left( \frac{2\pi}{a} y \right) \bm{\hat{z}} + \omega \sin \left( \frac{2\pi}{a} z \right) \bm{\hat{x}} \\
\psi^2_{H_3} =  \cos \left( \frac{2\pi}{a}y \right) \bm{\hat{x}} + \omega^2 \cos \left( \frac{2\pi}{a} x \right) \bm{\hat{z}} +\omega\cos \left( \frac{2\pi}{a} z \right) \bm{\hat{y}}
\end{matrix}
\\ \hline
H_4 & 
\begin{matrix}
\psi_{H_4}^1 = \cos \left( \frac{2\pi}{a} z \right) \bm{\hat{y}} , & \psi_{H_4}^2 = \cos \left( \frac{2\pi}{a} x \right) \bm{\hat{z}} \\
\psi_{H_4}^3 = \cos \left( \frac{2\pi}{a} y \right) \bm{\hat{x}} , & \psi_{H_4}^4 = \sin \left( \frac{2\pi}{a} z\right) \bm{\hat{x}} \\
\psi_{H_4}^5= \sin \left( \frac{2\pi}{a} y \right) \bm{\hat{z}} , & \psi_{H_4}^6 = -\sin \left( \frac{2\pi}{a} x \right) \bm{\hat{y}}
\end{matrix}
\\ \hline \hline
\end{array}
$$
\label{table:basisH0}
\end{table}

As an example, let us consider the six $H$ points $\bm{k} = \pm b \hat{\bm{i}}\ (i=x,y,z)$, $b=2\pi/a$,  and we call them $H^{(0)}$ points.
They are the smallest wavevectors among the $H$ points in $k$-space.
At these points the basis functions are formed by the 12 plane waves in free space for the six wavevectors 
$\bm{k} = \pm b \hat{\bm{i}}\ (i=x,y,z)$ having two polarizations each. 
These basis functions are decomposed into irreps as summarized in Table \ref{table:basisH0}, by using the character table for the irreps of the $H$ point summarized in Table~S3 in Supplemental Document. Therefore,
the eigenmodes with the lowest frequency at the $H^{(0)}$ points are decomposed into irreps $H_1 + H_2H_3 + H_4$, where $H_2$ and $H_3$ are degenerate in the presence of TRS.

\subsection{Representations at the high-symmetry points in the BPI and DG photonic crystals}
\begin{figure}
\centering
\includegraphics[scale=0.28]{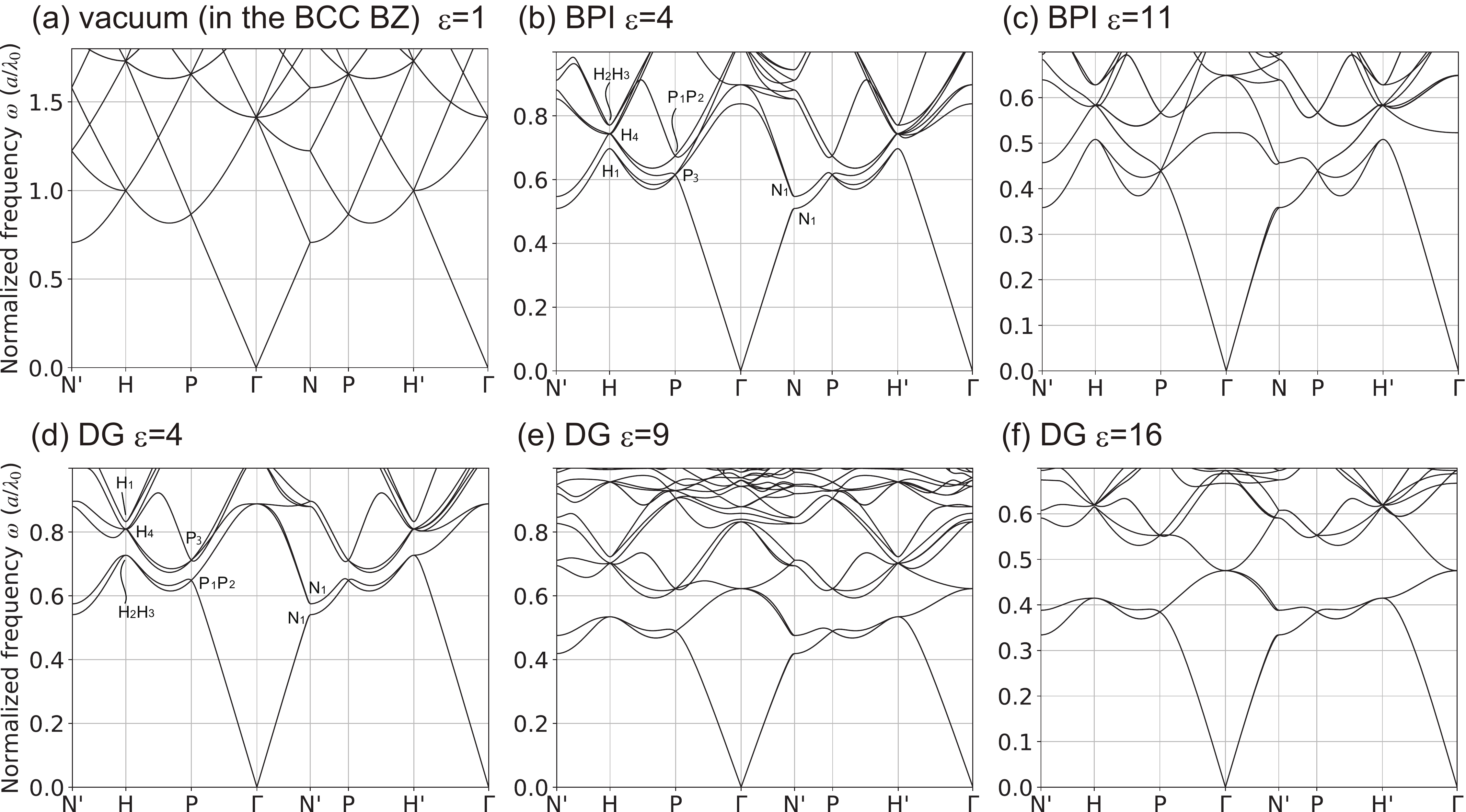}
\caption{Band structures for BPI and DG photonic crystals.
Band structures for (a) $\varepsilon=1$, which is an air band structure embedded into the Brillouin zone of the body-centered cubic (BCC) photonic crystal. Band structures are shown for the BPI photonic crystal (b) at $\varepsilon=4$, (c) at $\varepsilon=11$, and for the DG photonic crystal (d) at $\varepsilon=4$, (e) at $\varepsilon=9$, (f) at $\varepsilon=16$. In (b) and (d) we show the irreps for some of the lowest bands at $H$, $P$ and $N$ points mentioned in the text. There is (a) a twelve-fold, (b), (c) two-fold ($H_1$ irrep), and (d)-(f) four-fold degeneracy ($H_2H_3$ irreps) in the lowest bands at the $H$ point.}
\label{fig:phc_evolution}
\end{figure}

This result based on the plane waves matches with our numerical calculations for the BPI and the DG photonic crystals with various values of the dielectric constant $\varepsilon$, as shown
in Fig.~\ref{fig:phc_evolution}. The lowest four modes for the vacuum in Fig.~\ref{fig:phc_evolution}(a) splits into  $H_1 + H_2H_3 + H_4$  in 
both the BPI and the DG photonic crystals with increasing the dielectric constants of the dielectrics.
In the same manner,
one can classify the irreps for the several lowest frequencies at the high-symmetry points such as $H$,  $P$ and $N$.
These results are summarized in Table~S1 in Supplemental Document.

Let us discuss the change of the band structure and the irreps in the BPI photonic crystal, when the dielectric constant $\varepsilon$ in photonic crystal is varied from 1 (Fig.~\ref{fig:phc_evolution}(a)) via 4 (Fig.~\ref{fig:phc_evolution}(b)) to 11 (Fig.~\ref{fig:phc_evolution}(c)).
The degeneracies at the high-symmetry points such as $H$,  $N$ and $P$ at $\varepsilon=1$ are split into irreps in Table~S1 when $\varepsilon$ is increased.
In this case,
the irreps of the lowest bands at high-symmetry points are given by
$H_1$, $N_1$, and $P_3$ summarized in Table.~\ref{table:irreps_phc}.
These irreps remain the lowest bands even up to $\varepsilon=11$.

On the other hand, behaviors of the DG photonic crystal are different from that in the BPI photonic crystal.  The DG photonic crystal is shown
in Fig.~\ref{fig:phc_uc}(b) whose space group is $\bm{{\it 230}}$, being the same as the BPI photonic crystal.
The single gyroid surface is given by an isosurface of
\begin{align}
&g(\bm{r}) = \sin (2\pi x/a) \cos (2\pi y/a) + \sin (2\pi y/a) \cos (2\pi z/a) \nonumber 
\\
&+ \sin (2\pi z/a) \cos (2\pi x/a),
\label{eq:dg_function}
\end{align}
where $a$ is the lattice constant in the BCC lattice \cite{Wohlgemuth2001gyroid}, and the DG surfaces are given by isosurfaces of $g(\bm{r})$ and of its counterpart by space inversion, $g(-\bm{r})$. 
We set the dielectric regions as $g(\bm{r})> \lambda_{\mathrm{iso}}$ and $g(-\bm{r}) > \lambda_{\mathrm{iso}}$ with $\lambda_{\mathrm{iso}} = 1.1$, shown in blue and red, respectively in Fig.~\ref{fig:phc_uc}(b).
From our results on the band structures when the dielectric constant $\varepsilon$ is changed from 1 (Fig.~\ref{fig:phc_evolution}(a)) via 4 (Fig.~\ref{fig:phc_evolution}(d)) and 9 (Fig.~\ref{fig:phc_evolution}(e)) to 16 (Fig.~\ref{fig:phc_evolution}(f)), the irreps of the four lowest bands at high-symmetry points are given by $H_2H_3$, $N_1$, and $P_1P_2$ which are summarized in Table~\ref{table:irreps_phc}.
This set of the irreps are different from the BPI photonic crystals.

\begin{table}
\caption{A set of irreducible representations at high-symmetry points $H$, $N$, and $P$ for the lowest bands in BPI and DG photonic crystals.
The numbers in the parenthesis show the dimensions of the irreps.}
$$
\begin{array}{cccc} \hline \hline
\mathrm{high\mbox{-}symmetry \ point} & {H} & {P} & {N} \\ \hline
\mathrm{BPI \ photonic \ crystal} & H_1(2) & P_3(4) & N_1(2) \\
\mathrm{DG \ photonic \ crystal} & H_2H_3(4) & P_1P_2(4) & N_1(2) \\ \hline \hline
\end{array}
$$
\label{table:irreps_phc}
\end{table}

\subsection{Photonic waves at the $\Gamma$ point in photonic bands}
We have been focusing on the glide-$Z_2$ invariant for the 
lowest two bands of this photonic crystal.
Here, although the electromagnetic waves are singular at $\omega =0=|\bm{k}|$ as mentioned in Ref.~\cite{Watanabe2018prl121}, 
one can identify their symmetry properties near $\omega =0=|\bm{k}|$, which are needed for calculation of 
the $Z_2$ topological invariant.
First, the electromagnetic waves in a homogeneous isotropic medium are written analytically. 
Then, as remarked in Ref.~\cite{Watanabe2018prl121}, even in photonic crystals, the electromagnetic waves near $\omega =0=|\bm{k}|$
behave similarly with those in a homogeneous isotropic medium, because in the longwavelength limit, the 
spatial variation of the dielectric constants in the photonic crystal will become irrelevant.
In the next subsection, we show how we calculate the symmetry property of the waves around the singularity at $\omega=0$ within our theory.

\subsection{Glide-$Z_2$ invariant for the photonic crystal}
\label{sec:glideZ2}

In magnetic glide-symmetric systems, 
the glide-$Z_2$ topological invariant can be defined, which characterizes a topological crystalline insulator phase \cite{Fang2015prb91, Shiozaki2015prb91,Kim2019prb100}.
In Ref.~\cite{Lu2016nphys12}, the BPI photonic crystal is proposed to be in the topological phase ensured by this glide-$Z_2$ topological invariant, 
but a physical reason why this structure leads to the topological phase is unknown. Here we show this reason by directly studying
the formula of the glide-$Z_2$ topological invariant.

\begin{figure}
\centering
\includegraphics[scale=0.29]{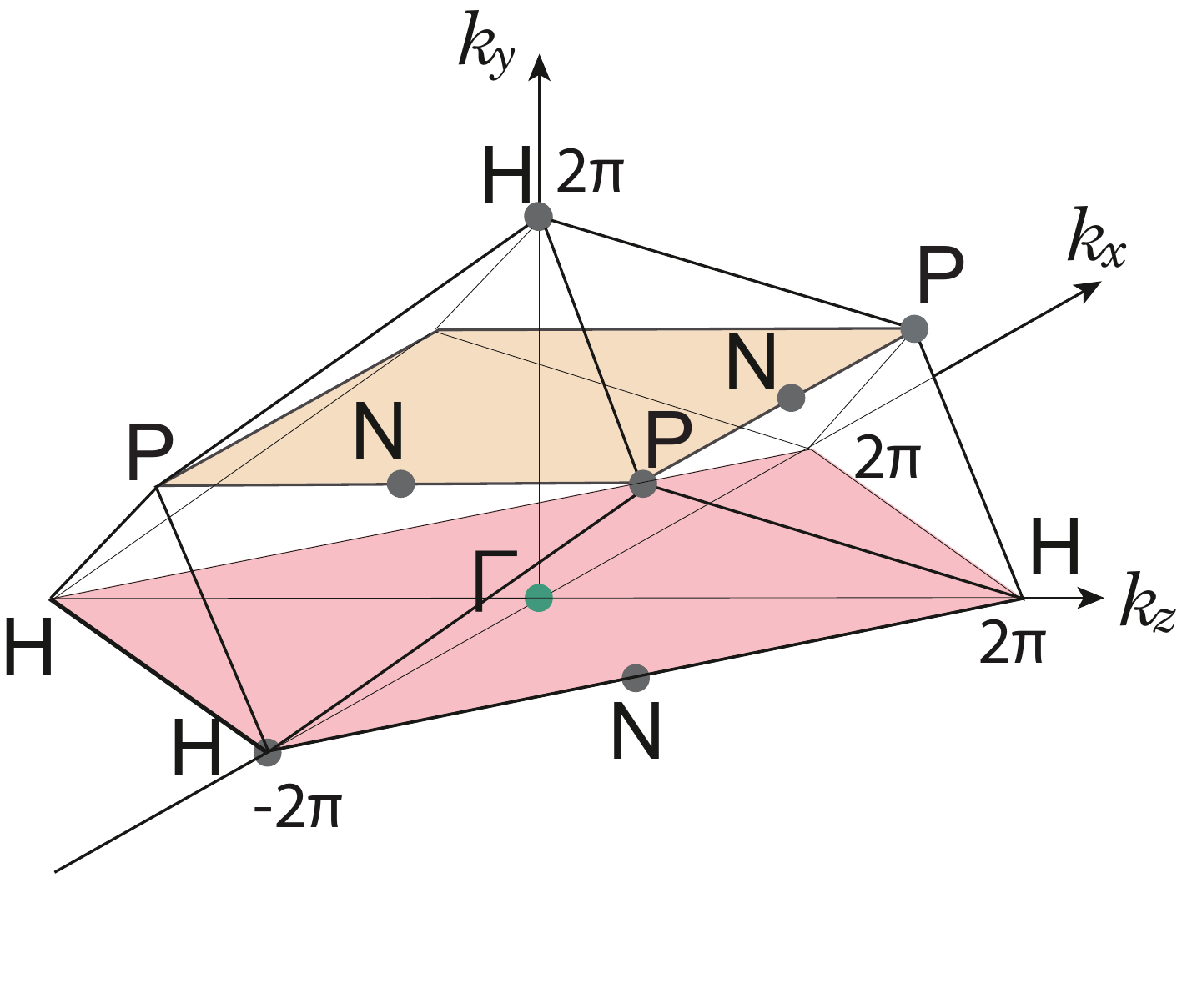}
\caption{(Color online) Upper half of the Brillouin zone of the monoclinic base-centered lattice in $\bm{{\it 230}}$ ($Ia\bar{3}d$). $\Gamma$, $P$,  $H$,  and $N$ denote the high-symmetry points in SG $\bm{{\it 230}}$. Here, all the lattice constants are set to be unity.}
\label{fig:bz7-15}
\end{figure}

The glide-$Z_2$ topological invariant $\nu$ is defined in a crystal with glide symmetry, 
and its formula is expressed as a sum of line and surface integrals in $k$-space \cite{Fang2015prb91, Shiozaki2015prb91,Kim2019prb100},
which is not easy to evaluate numerically.
On the other hand, when inversion symmetry is preserved, like the photonic crystals of our interest, its formula is
drastically simplified, and it is expressed in terms of the irreps at high-symmetry points\cite{kim2020-nonprimitive}
in the form of a symmetry-based indicator
\cite{Po2017ncommun8,
Bradlyn2017nature547}; thus it is much easier to evaluate. 
In the space group $\bm{{\it 230}}$, which our photonic crystals belong to, 
we obtain a formula for the glide-$Z_2$ topological invariant $\nu$, associated with a certain band gap, as
\begin{align}
&
(-1)^\nu = \prod_{i} \zeta^+_i (\Gamma) \xi_i (N) \frac{\xi_i (H)}{\zeta_i^+ (H)},
\label{eq:z2-230}
\end{align}
where the product is taken over the bands below the gap considered, $\zeta_i^+$ is a $C_2$ eigenvalue in the glide 
sector with a glide eigenvalue $g_+=e^{-ik_z/2}$, $\xi_i$ is an inversion parity for the $i$-th occupied state, and the product runs over 
the bands below the gap considered. The detailed derivation of this formula is presented in Section S2 in the Supplemental Document.
The value of $\nu$ is either $\nu=0$ or $\nu=1$, corresponding to the topologically trivial and nontrivial phases, respectively. 
The high-symmetry points in the Brillouin zones for $\bm{{\it 230}}$ are shown in Fig.~\ref{fig:bz7-15}. 
We note that from Eq.~(\ref{eq:z2-230}), the irreps at the $P$ point in $\bm{{\it 230}}$ do not contribute to the glide-$Z_2$ invariant, as we mentioned in the last part of the 
Section 2.

Remarkably, in $\bm{{\it 230}}$, the formula (\ref{eq:z2-230}) for the $Z_2$ topological invariant is further simplified 
drastically, by evaluating Eq.~(\ref{eq:z2-230}) term by term. 
In this calculation, we focus on the band gap between the second and third bands, when TRS is broken by applying a magnetic field.
First, the $H$ point has three possible physically irreducible representations, 
$H_1$, $H_2H_3$, and $H_4$, and all of them yield $\prod_{i} \frac{\xi_i (H)}{\zeta_i^+ (H)}=1$, by a direct calculation
from Table~S4 in Supplemental Document.
Second, the $N$ point has two possible physically irreducible representations, $N_1$ and $N_2$. Both of them contributreare 
to the product in 
Eq.~(\ref{eq:z2-230}) by a factor
$\prod_{i\in N_{a}} \xi_i (N)=-1$ $(a=1,2)$. Therefore, the total contribution from the $N$ point is $(-1)^{n/2}$, where $n$ is the number of bands below the gap considered.
Third, as noted earlier, the eigenmodes with $\omega=0$ at the $\Gamma$ point are
special in any photonic crystals because of the transversality condition of the electromagnetic wave,
and they do not follow the ten irreps at the $\Gamma$ point. 
Through a direct calculation, we show that it contributes trivially (i.e. by a factor $+1$) to the product in Eq.~(\ref{eq:z2-230}).
To summarize, 
the glide-$Z_2$ invariant $\nu$ for $\bm{{\it 230}}$ is calculated as \begin{align}
&(-1)^\nu =(-1)^{n/2}\ \rightarrow\ \nu=n/2\ (\text{mod}\ 2).
\label{eq:z2-230-2}
\end{align}
It means that when the number of bands $n$ below the gap is $n=4m+2$ ($m$: integer), 
the photonic crystal is topologically nontrivial, and when it is $n=4m$ ($m$: integer), it is topologically trivial, provided the gap is open everywhere in the Brillouin zone.

At the $P$ point, there are two possible physically irreps under the TRS, $P_1P_2$ and $P_3$, both of which are 
four-dimensional. Therefore, when $n=4m+2$ (i.e. $\nu=1$), there is no gap at the $P$ point when the TRS is preserved. Therefore we need 
to open a gap at $P$ by breaking TRS to make it topologically nontrivial. On the other hand, when $n=4m$ (i.e. $\nu=0$), the $P$ point is
gapped. 
Here, the $P_1P_2$ and $P_3$ irreps lead to the doubly degenerate Dirac point, and the generalized Dirac point, 
respectively, and in both cases, quad-helicoid surface states are realized on the (001) surface, 
thanks to the combination of the two glides and the
time-reversal symmetries
\cite{CLEO,PhysRevLett.124.104301}, and another $Z_2$ topological invariant, which we call 
a helicoid-$Z_2$ invariant, could be defined to represent nontrivial winding of the helicoid surface states. 
How the helicoid-$Z_2$ and glide-$Z_2$ are mutually related is an open question.%

Thus far, we have understood how to make the glide-$Z_2$ topological invariant $\nu$ nontrivial.
To make the photonic crystal topological, it is important how we break the TRS, i.e. how we introduce magnetization into dielectrics.
Here we introduce two types of magnetizations employed in the previous works, and we call these types I and II. 
The type I is realized in 
the BPI photonic crystal with 
the two dielectric rods (red and yellow in Fig.~\ref{fig:phc_uc}) 
having $+z$ magnetization, and the other two (blue and green in Fig.~\ref{fig:phc_uc})
having $-z$ magnetization, as employed in 
Ref.~\cite{Lu2016nphys12}.
It is classified as the magnetic space group (MSG) $\bm{{\it 142.565}}$, 
whose list is in the Bilbao Crystallographic Server \cite{Aroyo2006bilbao}.
The type II is the DG photonic crystal with the uniform magnetization along the $+z$ direction employed in Ref.~\cite{Lu2013nphoton7},
which belongs to the MSG $\bm{{\it 142.567}}$.
As we have seen in this paper, 
the photonic crystal is in the topological phase, if a gap is open between the second and the third lowest bands.
From an argument based on band topology, we conclude that in the type I the photonic crystal opens a gap 
between the second and third lowest bands
everywhere in $k$-space, while in the type II, the photonic crystal does not open a gap. This is explained by extending the notion of minimal band 
connectivity $M$ for a nonmagnetic 
photonic crystal in Ref.~\cite{Watanabe2018prl121} into magnetic photonic crystals, whose details are
shown in Section S2 in the Supplemental Document. This conclusion fully agrees with the results 
in Refs.~\cite{Lu2016nphys12,Lu2013nphoton7}, saying that the BPI photonic crystal is in a topological insulator phase ensured 
by glide symmetry for the gap between the second and the third bands, while the DG photonic crystal does not open a gap.

We note that in general, some other symmetries apart from glide symmetries can give topological surface modes. For example, 
mirror symmetry can give us topological phases associated with a mirror Chern number, leading to 
topological surface modes in the gap
\cite{PhysRevB.78.045426}.
Meanwhile, 
the space group $\bm{{\it 230}}$, which we study in this paper, contains various symmetries such as threefold rotations and twofold screw rotations other than glide symmetries, 
but
none of the other symmetries except for glide symmetries
are associated with topological surface modes. 
Thus in this space group $\bm{{\it 230}}$, the band gap topology solely depends on the glide symmetry.

\section{Design of topological photonic crystals with glide symmetry}
\label{sec:manipulation}
So far we have seen how the topological properties of these photonic bands are understood in terms of the irreps 
at high-symmetry points in $k$-space.
In this section,
we propose how topological photonic crystals are designed, based on the representation theory and Wyckoff positions.
In particular, we focus on the BPI and DG photonic crystals as two characterisitic examples, and how the photonic bands of
these photonic crystals result from their structure in real space. To show this we focus on the irreps at the $H$ point, 
and as we noted, in the BPI photonic crystal, the irrep $H_1$ is the lowest, while in the DG photonic crystal, the 
irreps $H_2H_3$ are the lowest.

\subsection{Perturbation theory}
\label{sec:perturbation}
As we argued in the previous section,
the irreps for the lowest bands at the $H$ point for the BPI photonic crystal and DG photonic crystal are unchanged by a change of the dielectric constant 
from unity to a larger value. We need to see how the level splitting occurs at the $H$ point by increasing $\varepsilon$.
In order to see the level splitting at the $H$ point for a value of $\varepsilon$ close to unity, we can employ the perturbation theory in the dielectric function.
When we add a small perturbation of the dielectric function $\varepsilon=1\ \rightarrow\ \varepsilon=1+\delta \varepsilon(\bm{r})$,
the frequency shift $\delta \omega$ is given to the first order in the perturbation $\delta \varepsilon(\bm{r})$ as follows \cite{Joannopoulos2008princeton}:
\begin{equation}
\delta \omega = - \frac{\omega}{2} \frac{\int d^3 \bm{r} \ \delta \varepsilon(\bm{r}) |\bm{\mathrm{E}}(\bm{r})|^2}{\int d^3 \bm{r} \ \varepsilon(\bm{r}) |\bm{\mathrm{E}}(\bm{r})|^2} .
\label{eq:perturb_phc}
\end{equation}
This relation implies that when the electric field is more concentrated within the dielectrics with having nonzero $\delta \varepsilon (\bm{r})$, the frequency shift increases.

We apply this formula to the lowest bands at the $H$ point. To this end we use the 12 plane wave basis functions at the lowest frequency at the $H^{(0)}$ points, decomposed into a set of eigenstates following the irreps $H_1$, $H_2$, $H_3$, and $H_4$, as summarized in Table~\ref{table:basisH0}.
Then, we can calculate the frequency shift by using Eq.~(\ref{eq:perturb_phc}),
if we know the spatial distribution of $\delta \varepsilon(\bm{r})$ corresponding to a photonic crystal.
As a result, within the first-order perturbation theory in Eq.~(\ref{eq:perturb_phc}),
one cannot determine the difference of the frequency shifts for the irreps $H_1, H_2$, and $H_3$ in Table~\ref{table:basisH0},
because the distribution of $|\bm{\mathrm{E}}|^2$ in Eq.~(\ref{eq:perturb_phc}) is identically the same for $H_1, H_2$, and $H_3$.
Here, the reason why the degeneracy is not lifted in the first order in $\delta \varepsilon(\bm{r})$ is because of the high symmetry of the cubic space group.
We expect that in the higher order in the perturbation $\delta\varepsilon(\bm{r})$, this degeneracy will be lifted, as required from 
the space-group symmetry.
In order to see how this degeneracy is lifted, we 
perform numerical calculations in the next subsection, 
instead of developing the higher-order perturbation theory, which is lengthy and complicated.

\subsection{Photonic band structures based on Wyckoff positions}
So far we have seen that in the present case it is not easy to see relationships between the structure of the photonic crystal and the band structure
in an analytic way. Therefore, we approach this problem with numerical calculations. Here, we 
introduce the notion of Wyckoff positions, often used in the context of electronic systems.
In electronic systems, Wyckoff positions classify spatial locations consistent with a given space group. By putting orbitals with various symmetry properties at  the Wyckoff positions, one can exhaust all the band structures of atomic insulators.
Inspired by this, we adopt the concept of the Wyckoff positions into our theoretical analysis on photonic crystals.

\begin{table}
\caption{Wyckoff positions in $\bm{{\it 230}}$, their site symmetries, and the irreducible representations (irreps) for the lowest bands with $\varepsilon = 12$ at the $H$ point, when dielectric spheres are put at the specified Wyckoff positions. If more than one bands have almost the same frequency, they are shown altogether as 
the lowest band.
Detailed coordinates of each Wyckoff position are summarized in Table~S5 in Supplemental Document.
The site symmetry $3$ for $32e$ is a subgroup of $\bar{3}$ for $16a$ and $32$ for $16b$, and the site symmetry $2$ for $48f$ and $48g$ is a subgroup of $32$ for $16b$, $222$ for $24c$, and $\bar{4}$ for $24d$.}
$$
\begin{array}{ccc} 
\hline \hline
\mathrm{Wyckoff \ positions} & \mathrm{Site \ symmetry} 
&\begin{matrix} \mathrm{Irreps \ for \ the} \\ \mathrm{lowest \ band \ at} \\ \mathrm{the \ } {H} \mathrm{\ point} \end{matrix}  \\[4pt]
 \hline
16a & \bar{3} (C_{3i}) & H_1,\ H_2H_3,\ H_4\\
16b &32 (D_3) &H_2H_3 \\
24c & 222 (D_2) &H_2H_3 \\
24d & \bar{4} (S_4) &H_1,\ H_2H_3 \\
32e & 3 (C_3) &  H_1 \\
48f & 2 (C_2) & H_1,\ H_2H_3 \\
48g&    2 (C_2) &   H_2H_3
\\ \hline \hline
\end{array}
$$
\label{table:230Wyckoff}
\end{table}

There are 8 kinds of Wyckoff positions in $\bm{{\it 230}}$ and their site symmetries are summarized in Table~\ref{table:230Wyckoff} \cite{Hahn2002ITA}.
Their positions for $16a$, $16b$ and $32e$ are in Table \ref{table:230WyckoffPosition}, and other Wyckoff positions are summarized in Table~S5 in Supplemental Document.
Our objective is to find which Wyckoff positions correspond to the band structure of the BPI and to that of the DG  photonic crystals.
To this end, we numerically calculate the band structure by putting dielectric spheres on a given Wyckoff position.
When dielectric spheres are absent, the light propagates in vacuum whose band structure is the same in Fig.~\ref{fig:phc_evolution}(a), but if the dielectric spheres have a finite radius, the band structure changes, depending on the positions of the dielectric spheres.

\begin{figure}
\centering
\includegraphics[scale=0.28]{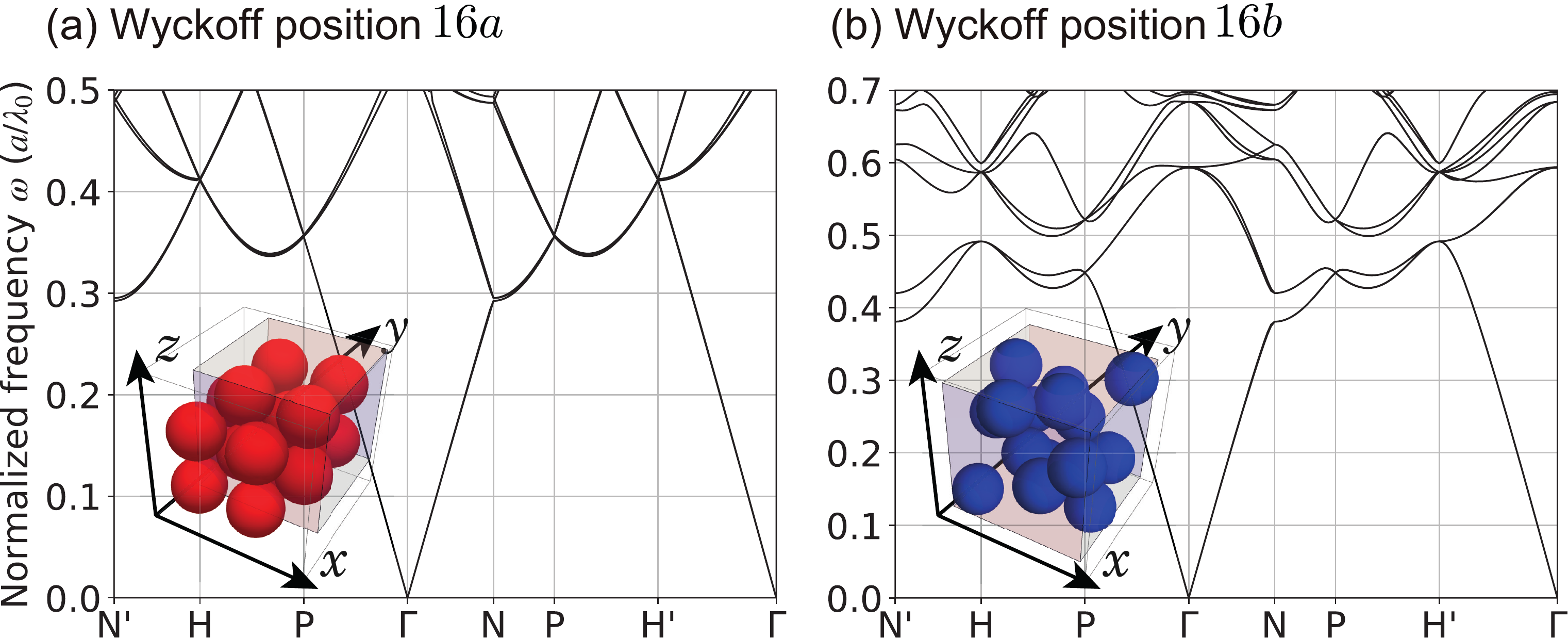}
\caption{The photonic band structures at $\varepsilon=12$ by putting dielectric spheres on the Wyckoff positions (a) $16a$ and (b) $16b$. 
(a) Band structure with dielectric spheres with the radius $\sqrt{3}/8$ at the Wyckoff position $16a$. It is hardly changed from an air band structure of a body-centered cubic photonic crystal. (b) Band structure with dielectric spheres with the radius $0.18$ at the Wyckoff position $16b$, not touching each other. The band structure is similar to that of the DG photonic crystal. 
Here we set the lattice constant to be unity.}
\label{fig:phc_abe}
\end{figure}

First, we put dielectric spheres at the Wyckoff positions labeled $16a$, which are located at 
$(\frac{n_1}{2}, \frac{n_2}{2}, \frac{n_3}{2} )$ and $(\frac{n_1}{2}+\frac{1}{4}, \frac{n_2}{2}+\frac{1}{4}, \frac{n_3}{2}+\frac{1}{4})$ with $n_1,n_2,n_3$ being integers,
according to Table~\ref{table:230WyckoffPosition}.
We here set the dielectric constant of the dielectric spheres to be $\varepsilon = 16$.
We set the radius of the dielectric spheres to be the maximum one, i.e the radius when they touch each other.
In this case, the band structure hardly changes from that for vacuum
as seen from Fig.~\ref{fig:phc_abe}(a).
Recall that Eq.~(\ref{eq:perturb_phc}) implies the degree of concentration of the electric fields in the dielectric regions.
Since the splitting of frequencies among the eigenmodes with $H_1, H_2H_3,$ and $H_4$ in the case of the Wyckoff position $16a$ (in Fig.~\ref{fig:phc_abe}(a)) are not conspicuous, the electromagnetic waves in these eigenmodes are localized on
the dielectric spheres at the Wyckoff position $16a$ to the same degree.

In the similar manner, we can calculate the corresponding band structures for dielectric spheres at other Wyckoff positions in $\bm{{\it 230}}$, listed in 
Table \ref{table:230Wyckoff}.
As a consequence, we find that dielectric spheres at the Wyckoff positions labeled by $16b, 24c$, and $48g$ generate band structure, with $H_2H_3$  being the lowest bands at $H$, similar to that from the DG photonic crystal, and the Wyckoff position labeled by $32e$ generates band structure with $H_1$ being the lowest bands at $H$, similar to that from the BPI photonic crystal. 
For the Wyckoff position $24d$, the eigenmodes for $H_1$ and $H_2H_3$ are also very close to each other, while they are away from those for $H_4$.
The lowest eight bands at the $N$ point and those at the $P$ point hardly split. 
For the Wyckoff position $48f$, the band structure hardly changes from vacuum.
We skip the most general Wyckoff position $96h$, because no special feature from symmetry is expected.
Remarkable cases are depicted in Fig.~\ref{fig:phc_abe}, where we set the dielectric constant as $\varepsilon = 12$ and the radius of dielectric spheres as $r = \sqrt{3}/8$ for $16a$ (Fig.~\ref{fig:phc_abe}(a)) and $r = 0.18$ for $16b$ (Fig.~\ref{fig:phc_abe}(b)) with the lattice constant being unity.

Let us compare these results with the BPI and DG photonic crystals in the related previous works. In the BPI photonic crystals, 
the dielectric rods lie along the $(111), (\bar{1}11), (1\bar{1}1),$ and $(11\bar{1})$ directions, which corresponds to 
the Wyckoff position $32e$, with changing the free parameter $x$ in Table \ref{table:230WyckoffPosition}. 
Indeed, the photonic crystal with dielectric spheres at  the Wyckoff position $32e$
was shown to have $H_1$ as the lowest band (see Table 3), similar to the BPI photonic crystal.

Next we discuss the DG photonic crystal with dielectrics located at $g(\bm{r})> \lambda_{\mathrm{iso}}$ and $g(-\bm{r}) > \lambda_{\mathrm{iso}}$ where $g(\bm{r})$ is defined in Eq.~(\ref{eq:dg_function}).
The isosurface function $g(\bm{r})$ has a maximum value $1.5$ when $\bm{r} = (x, y, z)$ is equal to the Wyckoff position $16b$.
By changing $\lambda_{\mathrm{iso}}$ smaller from $\lambda_{\mathrm{iso}} = 1.5$,
the dielectric regions for the DG photonic crystal broaden from the sites of $16b$ along $48g$ toward $24c$, i.e., 
along the planes perpendicular to the four $C_3$ rotation axes.
Indeed, in the band structures for $16b$, $48g$ and $24c$, the lowest bands follow the $H_2H_3$ irreps, and their band structures are simiar to that of the DG photonic crystal.

\begin{table}
\caption{Summary of Wyckoff positions $32e$, $16b$ and $16a$ in $\bm{{\it 230}}$. The first column denotes the multiplicity and the Wyckoff letter, the second column denotes site symmetry, and the coordinates with two sets $(0, 0, 0)+$ and $(\frac{1}{2}, \frac{1}{2}, \frac{1}{2})+$. The lattice constant is set to be unity.}
$$
\begin{array}{cc cccccccc}
\hline \hline
32e & . \ 3 \ . & \multicolumn{2}{c}{x,x,x} & \multicolumn{2}{c}{\bar{x}+\frac{1}{2}, \bar{x}, x+\frac{1}{2}} & \multicolumn{2}{c}{\bar{x}, x+\frac{1}{2}, \bar{x}+\frac{1}{2}} & \multicolumn{2}{c}{x+\frac{1}{2}, \bar{x}+\frac{1}{2}, \bar{x}} \\[4pt]
& & \multicolumn{2}{c}{x+\frac{3}{4}, x+\frac{1}{4}, \bar{x}+\frac{1}{4}} & \multicolumn{2}{c}{\bar{x}+\frac{3}{4}, \bar{x}+\frac{3}{4}, \bar{x}+\frac{3}{4}} & \multicolumn{2}{c}{x+\frac{1}{4}, \bar{x}+\frac{1}{4}, x+\frac{3}{4}} & \multicolumn{2}{c}{\bar{x}+\frac{1}{4}, x+\frac{3}{4}, x+\frac{1}{4}} \\[4pt]
& & \multicolumn{2}{c}{\bar{x},\bar{x},\bar{x}} & \multicolumn{2}{c}{x+\frac{1}{2}, x, \bar{x}+\frac{1}{2}} &\multicolumn{2}{c}{x, \bar{x}+\frac{1}{2}, x+\frac{1}{2}} & \multicolumn{2}{c}{\bar{x}+\frac{1}{2}, x+\frac{1}{2}, x} \\[4pt]
& & \multicolumn{2}{c}{\bar{x}+\frac{1}{4}, \bar{x}+\frac{3}{4}, x+\frac{3}{4}} & \multicolumn{2}{c}{x+\frac{1}{4}, x+\frac{1}{4}, x+\frac{1}{4}} & \multicolumn{2}{c}{\bar{x}+\frac{3}{4}, x+\frac{3}{4}, \bar{x}+\frac{1}{4}} & \multicolumn{2}{c}{x+\frac{3}{4}, \bar{x}+\frac{1}{4}, \bar{x}+\frac{3}{4}} \\[8pt]
16b & . \ 3 \ 2 & \frac{1}{8}, \frac{1}{8}, \frac{1}{8} & \frac{3}{8}, \frac{7}{8}, \frac{5}{8} & \frac{7}{8}, \frac{5}{8}, \frac{3}{8} & \frac{5}{8}, \frac{3}{8}, \frac{7}{8} & \frac{7}{8}, \frac{7}{8}, \frac{7}{8} & \frac{5}{8}, \frac{1}{8}, \frac{3}{8} & \frac{1}{8}, \frac{3}{8}, \frac{5}{8} & \frac{3}{8}, \frac{5}{8}, \frac{1}{8} \\[8pt]
16a & . \ \bar{3} \ . & 0, 0, 0 & \frac{1}{2}, 0, \frac{1}{2} & 0, \frac{1}{2}, \frac{1}{2} & \frac{1}{2}, \frac{1}{2}, 0 & \frac{3}{4}, \frac{1}{4}, \frac{1}{4} & \frac{3}{4}, \frac{3}{4}, \frac{3}{4} & \frac{1}{4}, \frac{1}{4}, \frac{3}{4} & \frac{1}{4}, \frac{3}{4}, \frac{1}{4} \\[4pt]
\hline \hline
\end{array}
$$
\label{table:230WyckoffPosition}
\end{table}

\subsection{Design of topological photonic crystals based on Wyckoff positions}

\begin{figure}
\centering
\includegraphics[scale=0.29]{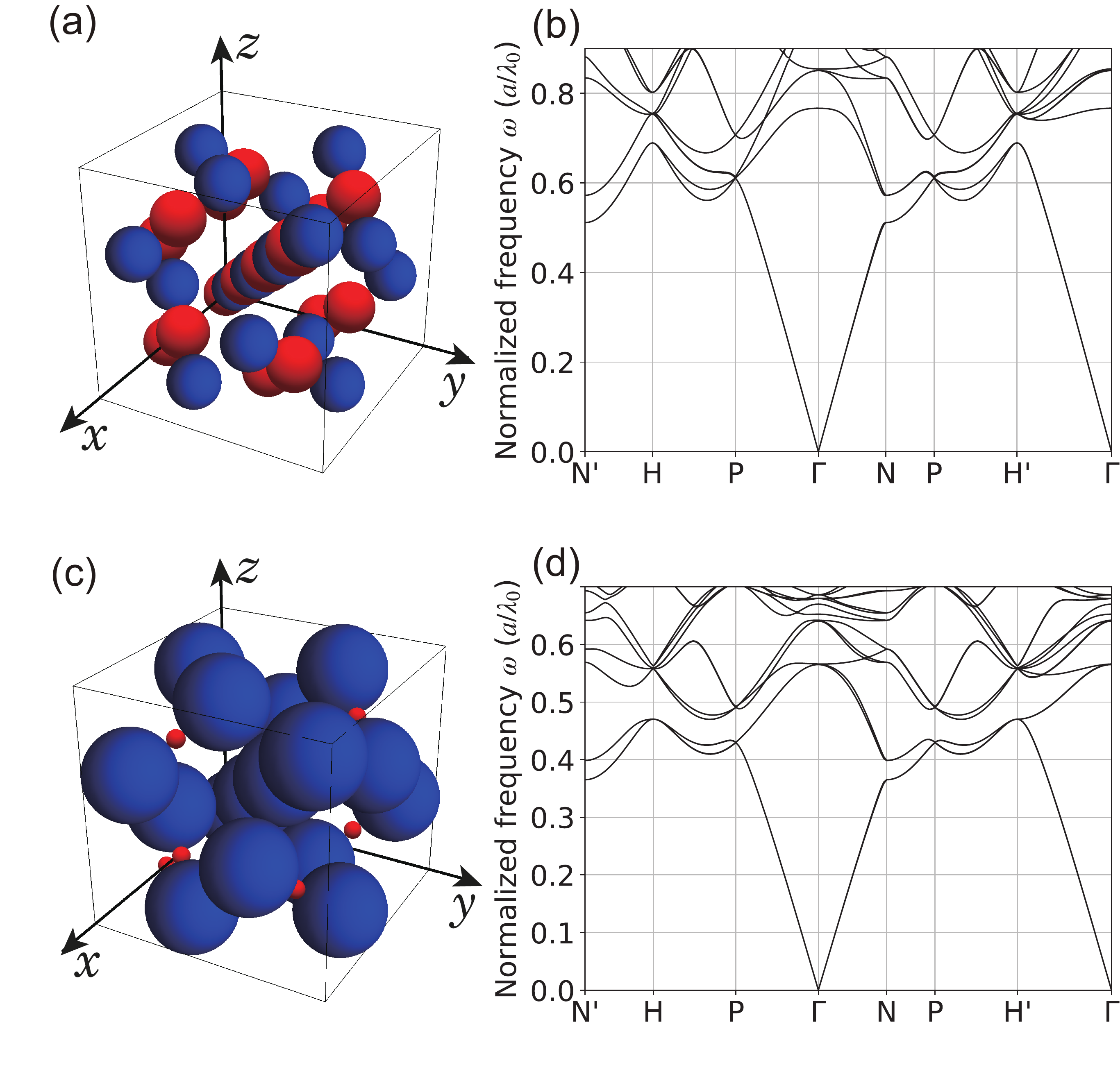}
\caption{Photonic crystals with dielectric spheres at the Wyckoff positions $16a$ and $16b$ and the corresponding band structures with time-reversal symmetry (TRS). 
(a) Photonic crystal with the dielectric spheres with the radius $\sqrt{3}/16$ at the Wyckoff positions $16a$ and $16b$. 
(b) The band structure is qualitatively similar to that of the BPI photonic crystal. (c) Photonic crystal with the dielectric spheres at the Wyckoff position $16b$ with the radii $5\sqrt{3}/48$ (blue) and those at the Wyckoff position $16a$ with the radii $\sqrt{3}/48$ (red). (d) The band structure is similar to that of the DG photonic crystal.}
\label{fig:phc_spheres}
\end{figure}

We have shown that dielectric spheres located at the Wyckoff position $32e$ in $\bm{{\it 230}}$ generates band structures, similar to the BPI photonic crystal, while those at $16b$ in $\bm{{\it 230}}$ generates band structures similar to the DG photonic crystal.
Now we discuss how these two cases are related if we gradually change the structure of the photonic crystal between the two cases. 
To this end,
 we focus on the Wyckoff positions, $16a$ and $16b$, which are regarded as special
 cases of the Wyckoff position 32$e$ by setting $x=0$ and $x=1/8$, repsectively (see Table \ref{table:230WyckoffPosition})

\begin{figure}
\centering
\includegraphics[scale=0.29]{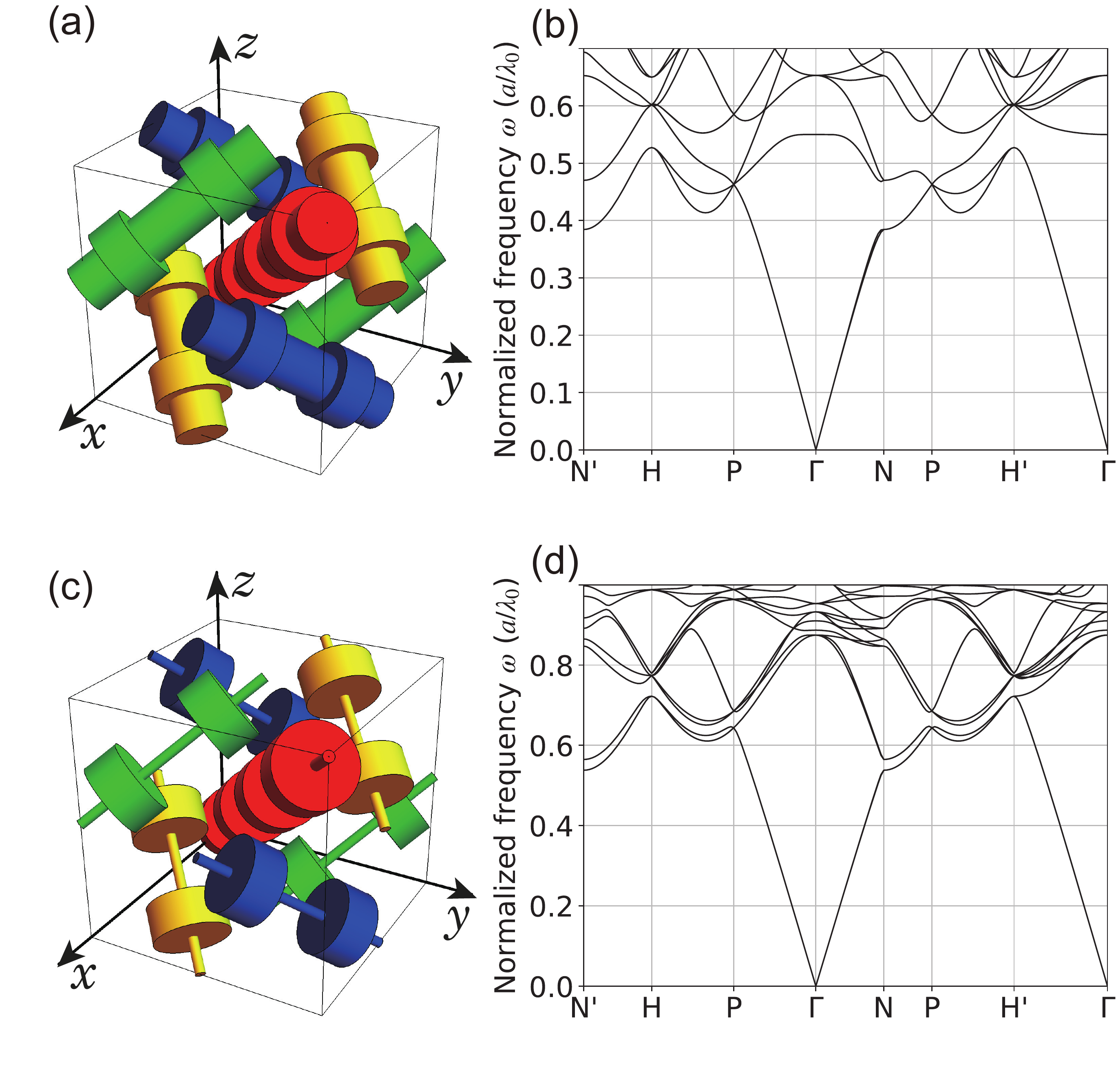}
\caption{Configurations of photonic crystals with dielectric cylinders and the corresponding band structures with time-reversal symmetry (TRS). (a) When the dielectric cylinders at the Wyckoff position $32e$ are dominant, the band structures in (b) is similar to those of the BPI photonic crystal. (c) If the dielectric cylinders at the Wyckoff position $16b$ are dominant, the band structures in (d) is similar to those of the DG photonic crystal.}
\label{fig:phc_cylinders}
\end{figure}

First, we examine band structures for photonic crystals with dielectric spheres both at $16a$ and at $16b$ for various values of their radii.
Let $r_a$ and $r_b$ denote the radii of dielectric spheres at the Wyckoff positions $16a$ and $16b$, respectively, and we change $r_a$ and $r_b$, with keeping the spheres touching each other, which leads to a constraint $r_a + r_b = \sqrt{3}/8$.
Henceforth, we set the lattice constant $a$ to be unity.
Let us start with $r_a = r_b = \sqrt{3}/16$. 
The configuration is depicted in Fig.~\ref{fig:phc_spheres}(a) where the gray cuboid represents a conventional unit cell $0\le x, y, z < 1$, and the band structure is shown in Figs.~\ref{fig:phc_spheres}(b).
The band structure  qualitatively agrees with that of the BPI photonic crystal, having the rods along the Wyckoff position $32e$(Fig.~\ref{fig:both_side}(b)), because the structure is similar to that of the BPI photonic crystal with dielectrics at $32e$, as can be seen 
in from Fig.~\ref{fig:phc_spheres}(a). 
Next, we make the system similar to the photonic crystal with spheres at $16b$, by putting $r_a = \sqrt{3}/48$ and $r_b = 5\sqrt{3}/48$ depicted in Fig.~\ref{fig:phc_spheres}(c). Then the band structure is given in Figs.~\ref{fig:phc_spheres}(d).
In this case, the band structure qualitatively agrees with that of the DG photonic crystal
having dielectrics mainly at 16$b$.

As another example, we also consider photonic crystals with dielectric cylinders, in order to see how the band structures changes between the two cases.
We combine two species of cylinders, one along $32e$ and the other centered at $16b$ with sharing their axes in common, and we consider the dielectrics located inside the union of these sets of cylinders.
We fix the height and radius of dielectric cylinders at $16b$ as $h_b = \sqrt{3}/12$ and $r_b = 0.15$, and change the radius $r_e$ of dielectric cylinders with an infinite height along $32e$. 
In the case of $r_e = 0.1$ shown in Fig.~\ref{fig:phc_cylinders}(a), where the photonic crystal is almost identical with the BPI photonic crystal,
the band structure with TRS in Fig.~\ref{fig:phc_cylinders}(b) is similar to that of the BPI photonic crystal, as we intuitively expected, with the lowest bands identical with those for the BPI photonic crystal in Table \ref{table:irreps_phc}.
On the other hand, in the photonic crystal with $r_e=0.02$  in Fig.~\ref{fig:phc_cylinders}(c), where the dielectrics are localized around the Wyckoff position $16b$, the band structure is shown in Fig.~\ref{fig:phc_cylinders}(d), and is similar to that of the DG photonic crystal,
as listed in Table \ref{table:irreps_phc}. 
These results show that by classifying the photonic crystals in terms of the Wyckoff positions of the dielectrics, one can predict the ordering of irreps at high-symmetry points
(see Table \ref{table:230Wyckoff}). Thus, photonic crystals with dielectrics at various Wyckoff positions can serve as building blocks to find ones with desired band ordering at high-symmetry positions, and it can be a way to ``design'' photonic crystals with desired properties.

Now we discuss universality and limitation of the method of designing topological photonic crystals based on Wyckoff positions.
We consider that our analysis in terms of Wyckoff positions works well in the perturbative regime discussed in Sec.~\ref{sec:perturbation}, where the dielectric constant of the dielectrics is close to that of vacuum. Indeed, in the present paper, we show that the ordering of the irreps of the bands can be predicted from 
the Wyckoff positions of the dielectrics, as summarized in Table \ref{table:230Wyckoff}, and it universally holds as long as the dielectric constant is close to that of vacuum.
If the dielectric constant of the dielectrics become larger, various bands with different irreps can be inverted and the band ordering may differ from 
that in Table \ref{table:230Wyckoff}.  


\section{Conclusion and discussion}
\label{sec:conclusion}
In the present paper, by investigating the relationship between band structures and space group representations, we study how to manipulate topological photonic crystals ensured by glide symmetry.
In particular, for the BPI photonic crystal without time-reversal symmetry, which is proposed to realize the glide-$Z_2$ magnetic topological crystalline insulator phase in the previous paper, we explain the physical reason of the topological phase by using our new formula of the glide-$Z_2$ invariant in terms of irreducible representations. 
By comparing band structures and irreps at high-symmetry points in our numerical calculation, we have also figured out that the photonic crystal with $\bm{{\it 230}}$ realizes the glide-$Z_2$ magnetic topological phase only from symmetry considerations.
In such photonic crystal designed in this way, by opening the gap between the second and third bands by breaking time-reversal symmetry in a staggered way (type I in the main text), we always obtain the glide-$Z_2$ topological crystalline insulator.

\begin{figure}
\centering
\includegraphics[scale=0.45]{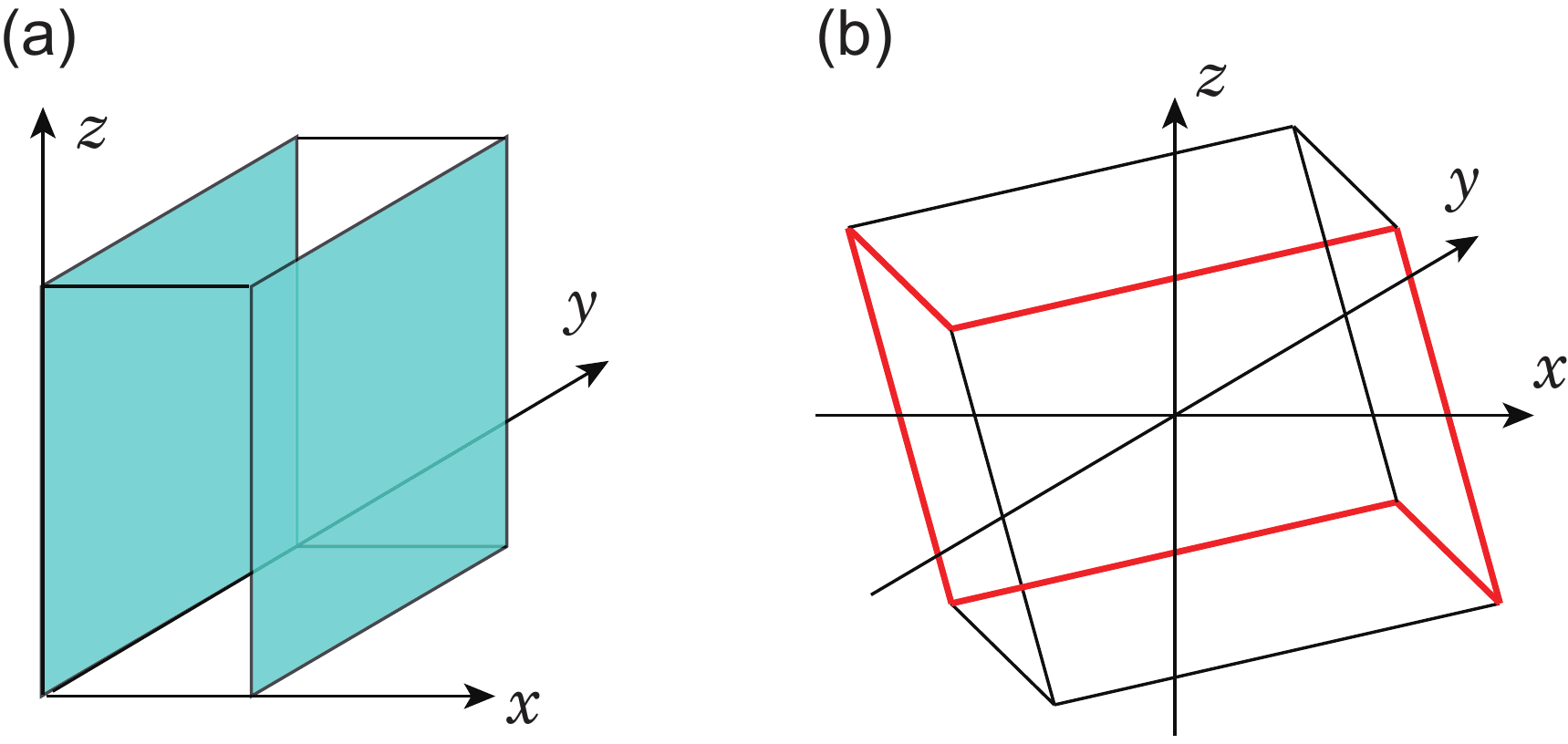}
\caption{Illustration of (a) the surface states (blue planes) of the glide-$Z_2$ topological crystalline insulators for system with glide-preserving surfaces, and (b) the hinge states (red lines) of the higher-order topological insulator ensured by inversion symmetry.}
\label{fig:surface_hinge_states}
\end{figure}

Moreover, because the glide-$Z_2$ topological crystalline insulator phase is automatically the higher-order topological insulator in the presence 
of inversion symmetry, such glide-$Z_2$ topological photonic crystals discussed in this paper are higher-order topological insulators, and 
one can expect topological hinge states \cite{Kim2019prb100}.
Under the glide operation $\hat{G}_y : (x,y,z) \rightarrow (x,-y,z+(c/2))$,
the $(100)$ surface preserves glide symmetry and the topological surface states of the glide-$Z_2$ topological phase emerge on this surface (Fig.~\ref{fig:surface_hinge_states}(a)).
On the other hand, if one consider a photonic crystal with an inversion-symmetric shape without glide-symmetric surfaces,
the topological surface states do not appear, while hinge states appear because inversion symmetry is still preserved (Fig.~\ref{fig:surface_hinge_states}(b))
\cite{Tanaka-hinge2020,PhysRevResearch.2.013300,tanaka2019appearance}.
This can be demonstrated by using a simple tight-binding model in Supplemental Document.
In electronic systems, it has been proposed that MnBi$_{2n}$Te$_{3n+1}$ and 
${\mathrm{EuIn}}_{2}{\mathrm{As}}_{2}$ support gapless surface states ensured by glide symmetry and hinge states ensured by inversion symmetry \cite{PhysRevLett.122.256402,PhysRevX.9.041039,Zhang2019arxiv1910}.

There remain several issues.
First, it is generaly difficult to open a common gap throughout the whole Brillouin zone, because
the gap induced by breaking time-reversal symmetry is small in general.
If the dielectric constant in the dielectrics is perturbatively introduced, $\varepsilon=1+\delta\varepsilon$, $\delta\varepsilon\ll 1$, 
the lowest frequency at the $H$ point is always higher than that at the $P$ point, because the $H$-$\Gamma$ distance is larger than the $P$-$\Gamma$ distance in $k$ space.
Even when the dielectric constant in the dielectrics becomes larger, the frequencies of the lowest bands at $P$ and $H$ are different, which means that a large TRS-breaking term is required to open a large gap.

In this paper, we focused on the topological photonic crystals protected by glide symmetry by breaking TRS. On the other hand, 
magnons in a magnet form band structure without TRS, and it can also be described by the same topological invariant. Therefore, we can use Eq.~(\ref{eq:z2-230}) to 
identify the glide-$Z_2$ topological invariant when inversion symmetry is present.
As an example, we can apply our theory to magnons in yttrium iron garnet (YIG), a ferromagnetic material belonging to the same space group $\bm{{\it 230}}$.
Nonetheless, so far the band structure of YIG has not been investigated enough for our purpose.
According to Ref.~\cite{Princep2017npjqm2}, the degree of degeneracy for the lowest bands at the $H$ point in the magnon band structure looks higher than four, which 
means that the irreps are not fully resolved, partially because of numerical difficulty due to the complex lattice structure having many atoms per unit cell.
Thus, topological characterization of magnons in YIG remains a future work.

\begin{backmatter}
\bmsection{Funding}
Japan Society for the Promotion of Science (JP17J10672, JP18H03678, JP20H04633); National Key R\&D Program of China (2017YFA0303800, 2016YFA0302400); Natural Science Foundation of China (12025409, 11721404, 11974415); Strategic Priority Research Program (XDB33000000); international partnership program (112111KYSB20200024); Beijing Natural Science Foundation (Z200008).

\bmsection{Acknowledgments}
H. K. is supported by Japan Society for the Promotion of Science (JSPS) KAKENHI Grant-in-Aid for JSPS Fellows Grant No.~JP17J10672.
This work was supported by JSPS KAKENHI Grant No. JP18H03678 and JP20H04633.
Ling Lu was 
supported by the National Key R\&D Program of China (2017YFA0303800, 2016YFA0302400), by Natural Science Foundation of China (12025409, 11721404, 11974415), by the Strategic Priority Research Program (XDB33000000) and the international partnership program (112111KYSB20200024) between the Chinese Academy of Sciences and the Croucher Foundation, and by Beijing Natural Science Foundation (Z200008).

\bmsection{Disclosures}
The authors declare no conflicts of interest.

\bmsection{Data availability statement}
 Data underlying the results presented in this paper are not publicly available at
this time but may be obtained from the authors upon reasonable request.

\bmsection{Supplemental document}
See Supplement 1 for supporting content.

\end{backmatter}


\end{document}